\documentclass[aps,prl,twocolumn,showpacs,groupedaddress]{revtex4-1}
\usepackage{graphicx}
\usepackage{dcolumn}
\usepackage{bm}

\begin{document}

\title{Optical measurement and modeling of interactions\\
between two hole or two electron spins in coupled InAs quantum dots}

\author{A. Greilich$^{1,2}$, \c{S}.~C. B\u{a}descu$^{1,3}$, D. Kim$^{1,4}$, A.~S. Bracker$^1$, D. Gammon$^1$}
\affiliation{
$^1$Naval Research Laboratory, Washington, DC 20375, USA\\
$^2$Technische Univerit\"at Dortmund, 44221 Dortmund, Germany\\
$^3$Air Force Research Laboratory, Wright-Patterson AFB, OH 45433, USA\\
$^4$Harvard University, Cambridge, MA 02138, USA}

\date{\today}

\begin{abstract}
Two electron spins in quantum dots coupled through coherent
tunneling are generally acknowledged to approximately obey
Heisenberg isotropic exchange. This has not been established for two
holes. Here we measure the spectra of two holes and of two electrons
in two vertically stacked self-assembled InAs quantum dots using
optical spectroscopy as a function of electric and magnetic fields.
We find that the exchange is approximately isotropic for both
systems, but that significant asymmetric contributions, arising from
spin-orbit and Zeeman interactions combined with spatial
asymmetries, are required to explain large anticrossings and
fine-structure energy splittings in the spectra. Asymmetric
contributions to the isotropic Hamiltonian for electrons are of the
order of a few percent while those for holes are an order of
magnitude larger.
\end{abstract}


\maketitle

The exchange interaction between the spins in two quantum dots (QDs)
leads to entanglement and to the opportunity of quantum information
processing~\cite{Loss1998}. Electron spin qubits have been studied
in great detail, but recently hole spins have received special
attention~\cite{DeGreeve2011,Greilich2011} because of their reduced
hyperfine interaction with the nuclear spin reservoir
~\cite{Brunner2009,Eble2009,Fallahi2010,Chekhovich2011}. Now, for
inter-QD entanglement, the nature of the exchange interaction is of
central importance, both for quantum gates and for decoherence of
two-qubit states~\cite{Burkard1999,Burkard2000,Dzhioev2002}.

The symmetry of the exchange interaction is often assumed to be the
Heisenberg isotropic exchange, $J{\bm\sigma}_1\cdot{\bm \sigma}_2$.
As a result, the eigenstates of two spins in two quantum dots form
singlet and triplet spin states separated by the exchange energy,
$J$. This model is very important, conceptually~\cite{Loss1998} and
for interpretations of complex experimental
spectra~\cite{Scheibner2007,Kim2011,Imamoglu12}, and has been used widely in magnetism,
quantum computing, molecular and quantum dot structures. It has been
shown that for two bound electrons the isotropic Heisenberg
interaction captures almost all of the physics and it requires only
small asymmetric exchange terms that couple singlets and
triplets~\cite{Kavokin2001,Badescu2005,Chutia2006,Baruffa2010a,Nowak2010,Stepanenko2012}.

In contrast to electron spins, hole spins are in some ways extremely
anisotropic; for example, in their $g$-factor~\cite{Andlauer2009}
and in their hyperfine interaction~\cite{Fisher2010}. This
anisotropy arises from the strong spin-orbit character of the
valence band, and is complicated by heavy-light hole mixing.
Counterintuitively, we find through optical spectroscopy that
isotropic exchange between two self-assembled InAs QDs is in fact a
good starting point for both two electrons and for two
holes~\cite{Kavokin2004}~\footnote{As a counter example,
electron-hole exchange in the case of the neutral exciton is nearly
Ising-like ($J\sigma_{ez}\sigma_{hz}$) with two nearly degenerate
doublets separated by the exchange splitting. In that case, a small
asymmetric $e$-$h$ exchange $\delta
(\sigma_x^h\sigma_y^e+\sigma_y^h\sigma_x^e)$ mixes the two bright
excitons. Another example is the case of two holes in a single QD --
one in the s-shell and the other in the p-shell. In contrast to two
electrons in a single QD, the energy structure deviates strongly
from isotropic, and is apparently neither Heisenberg-like nor
Ising-like~\cite{Ediger2007}.}. Nevertheless substantial asymmetric
contributions arising from spin-orbit interactions are necessary to
explain anti-crossings in the optical spectra. In addition,
inhomogeneous Zeeman interactions, that is, differences in the
$g$-factor between the two QDs and also in the tunnel barrier, lead
to additional energy splittings in the optical spectra that grow
with magnetic field. All of these interactions lead to off-diagonal
spin mixing terms that can be accounted for in a generalized spin
Hamiltonian.

\begin{figure}[t]
\includegraphics[width=\columnwidth]{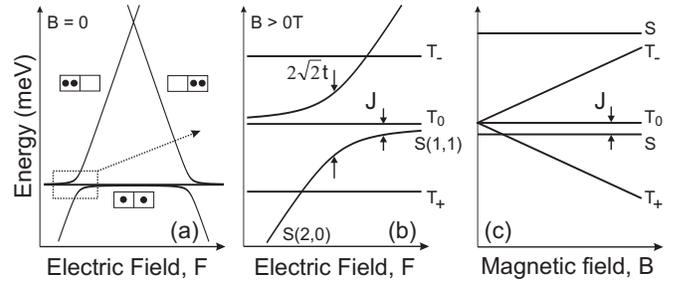}
\caption{\label{fig:motiv} (a) Hund-Mulliken model for two-spin
configuration at zero $B$. (b) $F$ dependence at fixed $B$ for a
symmetric case, Eq.~(\ref{eq:sym1}). (c) $B$ dependence at fixed $F$
(marked by arrows with $J$ in (b)).}
\end{figure}

In this paper we use individual pairs of vertically stacked
self-assembled InAs/GaAs QDs separated by a thin tunnel barrier with
a thickness $d$. Two types of samples were developed using a
Schottky diode grown by molecular beam epitaxy, one for
$2h$~\cite{Greilich2011} and one for $2e$~\cite{Kim2011}. To obtain
a direct comparison, the width and height of the tunnel barriers
were chosen to achieve similar values of singlet-triplet splitting
for both cases ($J\approx 100\,\mu$eV)~\cite{samples}. Electric
($F$) and magnetic ($B$) fields were applied longitudinally along
the stacking z-axis in the Faraday geometry. The optical spectra
were measured at 5\,K using photoluminescence with a spectral
resolution limited by the triple spectrometer of $\sim15\,\mu$eV for
the $2h$ case, and laser transmission spectroscopy with a resolution
of $<1\,\mu$eV for the $2e$ case.

The Heisenberg exchange can be treated within the Hund-Mulliken
model~\cite{vdWiel2002,Doty2008}. The natural spin state basis is
three triplets: $T_0=\left(\uparrow,\downarrow\right)_{T_0}$,
$T_+=\left(\uparrow,\uparrow\right)_{T_+}$,
$T_-=\left(\downarrow,\downarrow\right)_{T_-}$ and three singlets:
$S_{(2,0)}=\left(\uparrow\downarrow,0\right)_S$,
$S_{(1,1)}=\left(\uparrow,\downarrow\right)_S$, and
$S_{(0,2)}=\left(0,\uparrow\downarrow\right)_S$. The individual spin
projections are either the electron spin $\pm 1/2$ or hole
pseudo-spin $\pm 1/2$. The singlet states are coupled together and
shifted in energy by spin conserving tunneling ($t$) between the two
QDs, but because of spin blocking, the triplet states are not
affected. The Hamiltonian within the singlet basis is given by:
\begin{eqnarray}
\bordermatrix{~ & \left(\uparrow\downarrow,0\right)_S &
\left(\uparrow,\downarrow\right)_S &
\left(0,\uparrow\downarrow\right)_S\cr
      ~ & Fd + U & -\sqrt{2}t & 0 \cr
      ~ & -\sqrt{2}t & 0 & -\sqrt{2}t \cr
      ~ & 0 & -\sqrt{2}t & -Fd + U \cr}
\label{eq:sing}
\end{eqnarray}
The potential ($U$) is the Coulomb energy required to move the two
charges from separate QDs to the same QD. The relative energy
between QDs separated by $d$ is controlled via $F$. In
Fig.~\ref{fig:motiv}(a), the resulting levels of the singlets show
anticrossings. The lowest energy singlet state is $S = a S_{(2,0)} +
b S_{(1,1)} + c S_{(0,2)}$. In Fig.~\ref{fig:motiv}(b) and hereafter
we focus on one of the anticrossings, and take $c=0$. The isotropic
exchange interaction $J$ is defined as the splitting between $T_0$
and the lowest singlet with the spin Hamiltonian $J{\bm
\sigma}_1\cdot{\bm \sigma}_2$, with $J=J(F,U,t)$.

To fully probe and engineer the spin states of the two QDs we also
need, in addition to $F$, a magnetic field $B$. The simplest Hamiltonian
consists of isotropic exchange and an average Zeeman interaction,
\begin{equation}
J{\bm \sigma}_1\cdot{\bm \sigma}_2+{\bm \beta^{ext}}\cdot({\bm
\sigma}_1+{\bm \sigma}_2)\, \label{eq:sym1}
\end{equation}
The Zeeman term $\bm \beta^{ext}={(g_{11}+g_{22})\mu_B\bm B/2}$
splits the $T_+$ and $T_-$ lines, but preserves the spin states.
$\mu_B$ is the Bohr magneton. (1 and 2 mark the bottom and top dot,
respectively.) In Fig.~\ref{fig:motiv}(b) the energies are
calculated as a function of $F$ with $B$ held constant, while in
Fig.~\ref{fig:motiv}(c) the reverse is done.

\begin{figure}[t]
\includegraphics[width=\columnwidth]{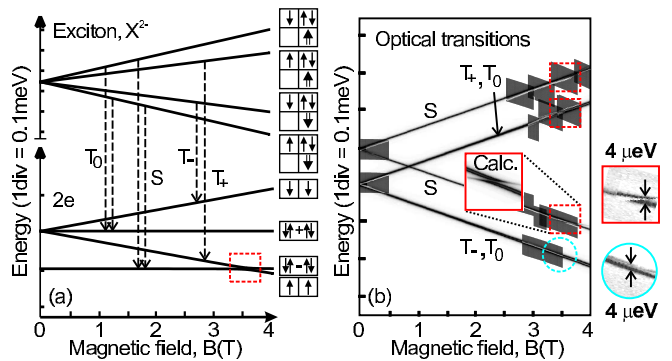}
\caption{\label{fig:symm} (color online) (a) Calculated $B$
dependence for the excited and ground states obtained by fitting the
$2e$ transmission spectra. Transitions are labeled by final state.
(b) Fitted optical spectrum for the $2e$ case using the parameters:
$g_{e,11} = g_{e,22} = -0.49$, $g_{h,22} = 1.58$, $t = 350\,\mu$eV,
$U=10$\,meV. Diamagnetic shift of 8.9\,$\mu$eV/T$^2$ is subtracted.
Shaded areas are the laser transmission measurements with the boxes
to the right representing an expanded views of the data at
$B=3.5$\,T.}
\end{figure}

The optical spectrum arises from transitions between the $2e$ levels
and the charged exciton levels ($X^{2-}$) as shown in
Fig.~\ref{fig:symm}(a). Here we show the calculated dependence on
$B$, holding $F$ constant. The exciton states have been described
previously~\cite{Scheibner2007,Doty2008}. The optical spectrum is
found by taking the difference between the $2e$ and ($X^{2-}$)
levels and weighting according to the selection rules. The results
for the $2e$'s, which are fits to the measured spectra, are shown in
Fig.~\ref{fig:symm}(b). As measured in previous studies, a singlet
line at $B=0$ splits into two with increasing $B$. Likewise, the
triplet line also splits into two, although each of these lines is
doubly degenerate. However, at high $B$ these lines also split,
although the splitting is very small for this $2e$ case and requires
the high resolution of laser spectroscopy to resolve. The lowest
inset on the right side of the figure shows the measured splitting
at $B=3.5$\,T with a value of 4\,$\mu$eV. The splitting grows to
8\,$\mu$eV at $B=5.6$\,T. We will show below that this
fine-structure splitting of the triplet arises from asymmetric
Zeeman terms. In addition there are small anticrossings that occur
only at $B=3.5$\,T as shown in the square insets on the right side
of Fig.~\ref{fig:symm}(b). These anticrossings occur where the $2e$
singlet and triplet energy levels would cross as seen in
Fig.~\ref{fig:symm}(a), and arise from asymmetric exchange due to
spin-orbit interactions. Other than these spin-flip anticrossings
and the triplet fine-structure splitting that grows with $B$, the
symmetric approximation is very good for the $2e$ spectrum.

\begin{figure}
\includegraphics[width=8cm]{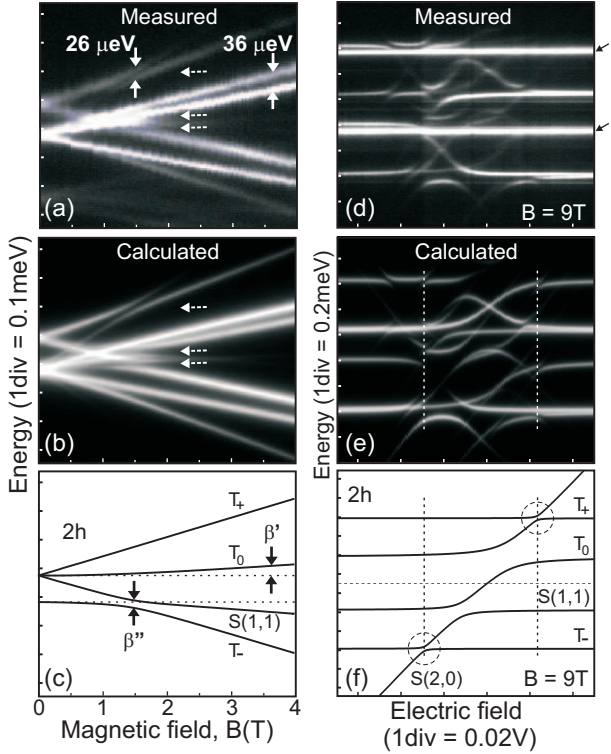}
\caption{\label{fig:hole} (a) measured and (b) calculated transition
spectra of $B$ dependence for the X$^{2+}$. $g_{h,11} =1.85$,
$g_{h,22}=0.78$, $g_{e,22} = -0.46$, $g_{h,12} = 0.14$, $t
=-115\,\mu$eV, $\gamma_{12} = 24\,\mu$eV, $\Delta \gamma
\sim5\,\mu$eV. In calculation
$S=\sqrt{0.44}S_{(2,0)}+\sqrt{0.56}S_{(1,1)}$. (a) and (b) Three
signatures of anticrossings are marked by dashed horizontal arrows.
Diamagnetic shift of 11.6\,$\mu$eV/T$^2$ is subtracted in (a). (c)
Calculated $2h$ energy levels with ground state $S/T_-$ anticrossing
of 26\,$\mu$eV. (d) measured and (e) calculated transition spectra
of $F$ dependence for $B=9$\,T. (small black arrows on the right of
panel (d) mark a transition spectra of X$^+$.) (f) Ground state
energy levels as a function of bias. Circles mark the spin-flip
anticrossings. Vertical lines mark the position of anticrossings in
transition spectra of Fig.~\ref{fig:hole}(d) and~\ref{fig:hole}(e).}
\end{figure}

In contrast to the $2e$ case of Fig.~\ref{fig:symm}(b), the measured
$2h$ spectrum appears much more complex as shown in
Fig.~\ref{fig:hole}(a). In addition to substantial fine-structure
splittings observed in the triplet optical transitions that grow
with increasing $B$ (36$\mu$eV at 3.5\,T), there are now obvious
anticrossings with magnitudes of 26\,$\mu$eV at $B=1.5$\,T that can
be observed even in photoluminescence, though with the resolution of
a triple spectrometer. We have also measured the $2h$ optical
spectrum as a function of $F$ at fixed $B$. The complex pattern of
anticrossings observed in the data of Fig.~\ref{fig:hole}(d) arises
from the spin-flip anticrossings in combination with the larger
spin-conserving anticrossings of both the $2h$ and the exciton
states that have been described previously~\cite{Doty2008,som}.
Remarkably, with the same formalism for both the $2e$ and the $2h$
cases, all of these features can be explained with natural
extensions of the symmetric Hamiltonian, Eq.~(\ref{eq:sym1}). As a
result we obtain the fitted optical transition spectrum shown in
Figs.~\ref{fig:hole}(b) and~\ref{fig:hole}(e), and calculate the
corresponding $2h$ energy levels in Figs.~\ref{fig:hole}(c)
and~\ref{fig:hole}(f). A comparison of these $2h$ energy levels and
those of the symmetric calculations of Fig.~\ref{fig:motiv} show
clearly the importance of asymmetric spin interactions.

We generalize the symmetric spin Hamiltonian of Eq.~(\ref{eq:sym1}) to
include spin coupling terms as follows. The spin interaction of
holes (or electrons) again is included with interaction terms, ${\bm
\beta}^{tot}_1\cdot{\bm \sigma}_1 + {\bm \beta}^{tot}_2\cdot{\bm
\sigma}_2$ acting on each spin. The fields now include not only the
external magnetic field but also the internal relativistic magnetic field
arising from the spin-orbit interaction $\bm \beta^{tot}_i=\bm
\beta^{ext}_i+\bm \beta^{so}_i$. We could also include the
Overhauser field due to the hyperfine interaction with nuclear
spins, in a similar way to Ref.~[\onlinecite{Stepanenko2012}], but
this is expected to be small here. After integrating over the
orbital degrees of freedom and including the isotropic exchange we
obtain the general spin Hamiltonian for two spins in two QDs:
\begin{equation}
J{\bm \sigma}_1\cdot{\bm \sigma}_2+{\bm \beta}\cdot({\bm
\sigma}_1+{\bm \sigma}_2)+{\bm \beta'}\cdot({\bm \sigma}_1-{\bm
\sigma}_2)+{\bm \beta''}\cdot({\bm \sigma}_1\times{\bm \sigma}_2)\,
\label{eq:mix-place1}
\end{equation}

In our case, the strong axis of the QDs and also the external $B$
are along the z-axis, ${\bm B}=B{\hat z}$. The QDs are also stacked
along the z-axis. We take the lateral asymmetry (e.g., an offset $d$
between the centers of the QDs) to have ${\bm
\beta}^{so}=\gamma{\hat x}$. This displacement generates  a gauge
factor $e^{i\phi_B}$ between the QDs, $\phi_B \propto Bd$
\cite{Baruffa2010b}. In Eq.~(\ref{eq:mix-place1}), ${\bm \beta}'=
Re\langle S|{\bm \beta}^{tot}|T\rangle $ and ${\bm \beta}''= Im
\langle S|{\bm \beta}^{tot}|T\rangle$ can be written in terms of the
Zeeman and spin-orbit fields ($\alpha= \mu_B B/2$):
\begin{eqnarray}
{\bm \beta}&=&\alpha\Sigma g\hat{\bm z}+\Sigma\gamma\hat{\bm x}\label{eq:h1}\\
{\bm \beta}'&=&\alpha (a g_{12}^{Re}+b\Delta g)\hat{\bm z}+(a
\gamma_{12}^{Re}+b\Delta \gamma)\hat{\bm x}\label{eq:h2}\\
{\bm \beta}''&=&\alpha a g_{12}^{Im}\hat{\bm z}+a
\gamma_{12}^{Im}\hat{\bm x}.\label{eq:h3}
\end{eqnarray}

The Zeeman interaction with the external $B$ is determined by the
matrix elements of the coordinate-dependent $g$-factor $g(r)$ over
the two QDs: $\Sigma g=(g_{11}+g_{22})$, $\Delta g=(g_{11}-g_{22})$
and $g_{12}=\langle 1|g(r)|2\rangle e^{i\phi_B}$. Likewise, the
spin-orbit field leads to mixing terms given by matrix elements of
$\gamma$. The linear superposition $aS_{(2,0)}+bS_{(1,1)}$ gives a
contribution of $\Delta \gamma$ and $\gamma_{12}$, which is
determined by measuring its dependence on $F$. The experiment gives
absolute values $|g_{12}|$ and $|\gamma_{12}|$.

We can also write the Hamiltonian of Eq.~(\ref{eq:mix-place1}) in a
matrix form within the singlet/triplet spin basis that shows
explicitly how the states are mixed, and permits convenient fitting
to the data.
\begin{equation}
\bordermatrix{~ & \left(\uparrow\downarrow,0\right)_S &
\left(\uparrow, \downarrow\right)_S &
\left(\uparrow,\downarrow\right)_{T_0} &
\left(\uparrow,\uparrow\right)_{T_+} &
\left(\downarrow,\downarrow\right)_{T_-}\cr
      ~& Fd + U & -\sqrt{2}t      & \alpha\sqrt{2}g_{12}& \gamma_{12}       & -\gamma_{12}      \cr
       & -\sqrt{2}t & 0 & \alpha\Delta g      & -\Delta\gamma & \Delta\gamma \cr
       & \alpha\sqrt{2}g_{12}^*& \alpha\Delta g & 0             &\Sigma\gamma & \Sigma\gamma \cr
       & \gamma_{12}^*     &-\Delta\gamma &\Sigma\gamma  & \alpha\Sigma g & 0 \cr
       & -\gamma_{12}^*    &\Delta\gamma & \Sigma\gamma  & 0 & -\alpha\Sigma g \cr}
\label{eq:asymm}
\end{equation}
Here we explicitly include both singlet states and the tunneling
rate $t$ (instead of $J$) for convenience in fitting the data.

The off-diagonal terms lead to the observed anticrossings and
fine-structure splittings in the spectra. The phenomenological
Zeeman and spin-orbit parameters, $g_{ij}$ and $\gamma_{ij}$, have
useful physical interpretations, and can each be associated with
specific features in the spectra. First, the term with $\Delta g$ is
manifesting the difference in $g$-factors between the two QDs, and
likely arises from the difference in size and indium concentration
between the two QDs. This parameter is larger for holes than for
electrons in part because the larger effective mass of the hole
makes them more localized and more sensitive to differences of the
QDs such as differences in size or composition. The term with
$g_{12}$ physically arises from the difference in $g$-factor between
the barrier and the QDs, which leads to the $\uparrow$ and
$\downarrow$ spins-tunneling at different
rates~\cite{Doty2006,Liu2011}. These terms breaks the spin symmetry
and mixes $S$ and $T_0$, effectively pushing the $S$ and $T_0$
energies apart as a function of $B$ as seen in
Fig.~\ref{fig:hole}(c). This is measured directly in the optical
spectrum by the fine-structure splitting in the triplet transitions
that grows with $B$. The splitting is given by $\beta'$ in
Eq.~(\ref{eq:h2}). At sufficiently high fields the eigenstates
become $(\uparrow,\downarrow)$ and $(\downarrow,\uparrow)$, instead
of $(\uparrow,\downarrow)_S$ and
$(\uparrow,\downarrow)_{T_0}$~\footnote{For holes, the difference in
$g$-factor of the InAs QDs and the GaAs tunnel barrier is
substantial and $g_{12}=0.14$~\cite{Doty2009}. For electrons it was
found that this difference was negligible with a GaAs barrier,
however, it has been found recently that with a GaAs/AlGaAs/GaAs
barrier, as used here, this term can also be significant
($g_{12}=0.3$)~\cite{Liu2011}. The symmetric Hamiltonian by itself
provides a good approximation to the $2e$ case.}.

The result of the spin-orbit field is similar but with an important
difference. The spin-orbit field acting orthogonal to
$\mathbf{\hat{z}}$ has the effect of partially rotating $\uparrow$
to $\downarrow$, and vice versa. In particular, the term
$\gamma_{12}$, which couples $(\uparrow\downarrow,0)_S
\leftrightarrow (\uparrow,\uparrow)_{T_-}$, can be viewed as
spin-flip tunneling. It is analogous to the spin conserving
tunneling term ($t$), which couples $(\uparrow\downarrow,0)_S
\leftrightarrow (\uparrow,\downarrow)_S$~\cite{Doty2010}. The
$\gamma_{12}$ term along with the $\Delta \gamma$ term have the
effect of mixing the $S$ state with the $T_-$ and $T_+$ triplet
states, and leads to the anticrossing observed in
Fig.~\ref{fig:hole}(c) with magnitude given by $\beta''$ in
Eq.~(\ref{eq:h3}).

The term $\Sigma \gamma$ couples $T_0$ with the $T_+$ and $T_-$
triplets. It can be measured as a splitting of the triplet line at
zero $B$. Any zero-field splitting of the triplet energies was found
to be less than our resolution, and so we took this parameter to be
zero~\footnote{We have set all $\gamma_{ii}$ terms to zero at $B\sim
0$ in Eq.~(\ref{eq:asymm}). $\Delta \gamma\gg \Sigma \gamma$
indicates that the local spin-orbit fields ${\bm \beta}^{so}_1$,
${\bm \beta}^{so}_2$ in the two QDs have opposite directions and
similar magnitudes. This does not allow for any splitting between
the $T_0$ and the $T_+ / T_-$ states, which must be less than our
resolution of 15\,$\mu$eV.}. We note that in a separate study there
is evidence for a finite splitting of $\sim 8\,\mu$eV for the $2h$
case in a similar sample from coherent measurements in the time
domain~\cite{Greilich2011}. Additionally, Ref.~[\onlinecite{som}]
represents a short description of a microscopic origin of these spin
mixing terms.

Using Eq.~(\ref{eq:asymm}) we are able to get good fits to the data
with the coupling parameters given in Table~\ref{tab:1}.

\begin{table}[h]
\caption{Electron and hole coupling parameters:}\label{tab:1}
\begin{ruledtabular}
\begin{tabular}{lcccccc}
& $\Sigma g$ & $\Delta g$ & $|g_{12}|$ & $\Sigma \gamma $($\mu$eV) & $\Delta \gamma$ ($\mu$eV)& $|\gamma_{12}|$ ($\mu$eV)\\
\hline
hole & 2.63 & 1.07 & 0.14 & 0 & 5 & 24\\
electron & 0.98 & 0 & 0.3 & 0 & 2 & 3\\
\end{tabular}
\end{ruledtabular}
\end{table}

In conclusion, we have found that a symmetric spin Hamiltonian based
on the isotropic Heisenberg exchange interaction can be generalized
to treat the $2h$ as well as the $2e$ spectrum in tunnel-coupled QDs
using phenomenological off-diagonal Zeeman and spin-orbit
parameters. The fact that the $2h$ exchange interaction between QDs
is primarily Heisenberg-like is important, because it means that
concepts and techniques developed for the control of  $2e$ spins can
potentially be used for $2h$'s. Moreover, the substantial spin
mixing that can occur at the anticrossing points and at large
magnetic fields is also potentially useful for spin control and/or
measurement. Such mixings have already been used to propose and
demonstrate simultaneous optical spin-flip and cycling transitions
for a single electron spin~\cite{Kim2008}. As another example,
quantum control of two holes could also be obtained with
electrostatic gates (instead of optical gates) in a way analogous to
Taylor~\textit{et al.}~\cite{Taylor_nucl}. However, instead of
hyperfine coupling, the much larger spin-orbit interaction could be
used, thereby enabling faster gates.

\bibliography{refs}

\end{document}